# SS Cygni: An analysis of quasi-periodic oscillations in the range of 0.25h to 8h.

Ian D. Sharp[1]

ian.sharp@astro-sharp.com

***Presented here is an analysis of over 66,000 magnitude measurements, made by the author, of the cataclysmic variable SS Cygni in two photometric filters (Cousins R and Johnson V). This large number of measurements, along with their cadence, is sufficient to analyse the light curve for quasi-periodic oscillations (QPOs) ranging from a few minutes to a few hours with the most frequently found QPOs having periods of 30 minutes (48 cycles/day). These QPOs occur predominantly during quiescent phases rather than outbursts, representing a previously understudied aspect of SS Cygni's variability.***

## Introduction

In 1896 Louisa D. Wells, a computer at the Harvard College Observatory, discovered the variability of SS Cygni (SS Cyg) and since that time it has become, perhaps, the most well studied variable star system of all. On a personal note, I have been observing this star since 1974 when I made a few visual magnitude estimates alongside my friend and mentor at that time: Dr John Mason (currently the director of the BAA Meteor Section). More recently, CCD photometry has become my main activity and I have acquired over 66,000 brightness measurements of SS Cyg since June 2024 and it is on that set of data this study is based.

SS Cyg is a cataclysmic variable (CV) star meaning one that irregularly increases in brightness by a large factor from a quiescent state. CVs are binary star systems that consist of a white-dwarf primary and a mass-transferring secondary which, in the case of SS Cyg is a red-dwarf with a temperature of 4560K. The binary period is approximately 0.275d (6.6h) and the semi-major axis separation is just over $1.8R_\odot$. SS Cyg shows no eclipse due to the binary inclination of ≈ 50°. More system parameters of SS Cyg can be found in **Voloshina et al. (2012).**

Figure 1 shows the light curve of SS Cyg using the magnitude measurements made by the author using Cousins R and Johnson V filters from 2024 June to 2026 January. The choice of R and V was made long before a large analysis was envisaged otherwise a photometric B filter might also have been chosen. However, it has been possible to keep the cadence of exposures higher with just R and V. This plot clearly shows the familiar irregular outburst in brightness which occur in the time interval of 25 to 73 days. Notice the peaks appear to come in wide and narrow varieties and that the magnitude typically varies between $V_{mag}$ = 12.0 at the dimmest parts of the quiescent phase and brightens to around $V_{mag}$ = 8.5 during outburst. The sample of outbursts shown in Figure 1 is typical of the hundreds of outbursts that have been recorded for well over 100 years by members of the British Astronomical Association (BAA) and the American Association of Variable Star Observers (AAVSO), however there was a hiatus in the usual pattern of outbursts around 2021 described by Jeremy Shears here on the BAA's YouTube channel in a BAA Weekly Webinar entitled "Is SS Cygni losing the plot?"

---

[1] British Astronomical Association, Ossington Chambers, 6/8 Castle Gate, Newark, NG24 1AX, UK

The instrumental magnitudes plotted in Figure 1 were obtained by performing ensemble differential aperture photometry using the AstroArt software package controlled by Python utilities written by the author. The images were captured using the author's equipment based in the UK and Spain. The telescope in the UK is a 235mm Schmidt-Cassegrain (SCT) and the one in Spain was a 280mm SCT. Both systems used a Starlight Xpress SX694 TRIUS PRO CCD camera and both operated at a focal ratio of f/7. However, the camera in Spain has recently been changed to a Moravian C3-26000 CMOS.

Notice in Figure 1 how the red and green points cluster together as vertical lines of varying height. This is because the dataset includes just over 300 nights of photometric measurements made up of many points per night. All exposures are 30 seconds and the average number of images per night is about 220 where the maximum number in any night was 1081. Because of this it is not possible to see any detailed structure in the light curve for any one night in Figure 1 due to the large time span on the x-axis covering about 600 days.

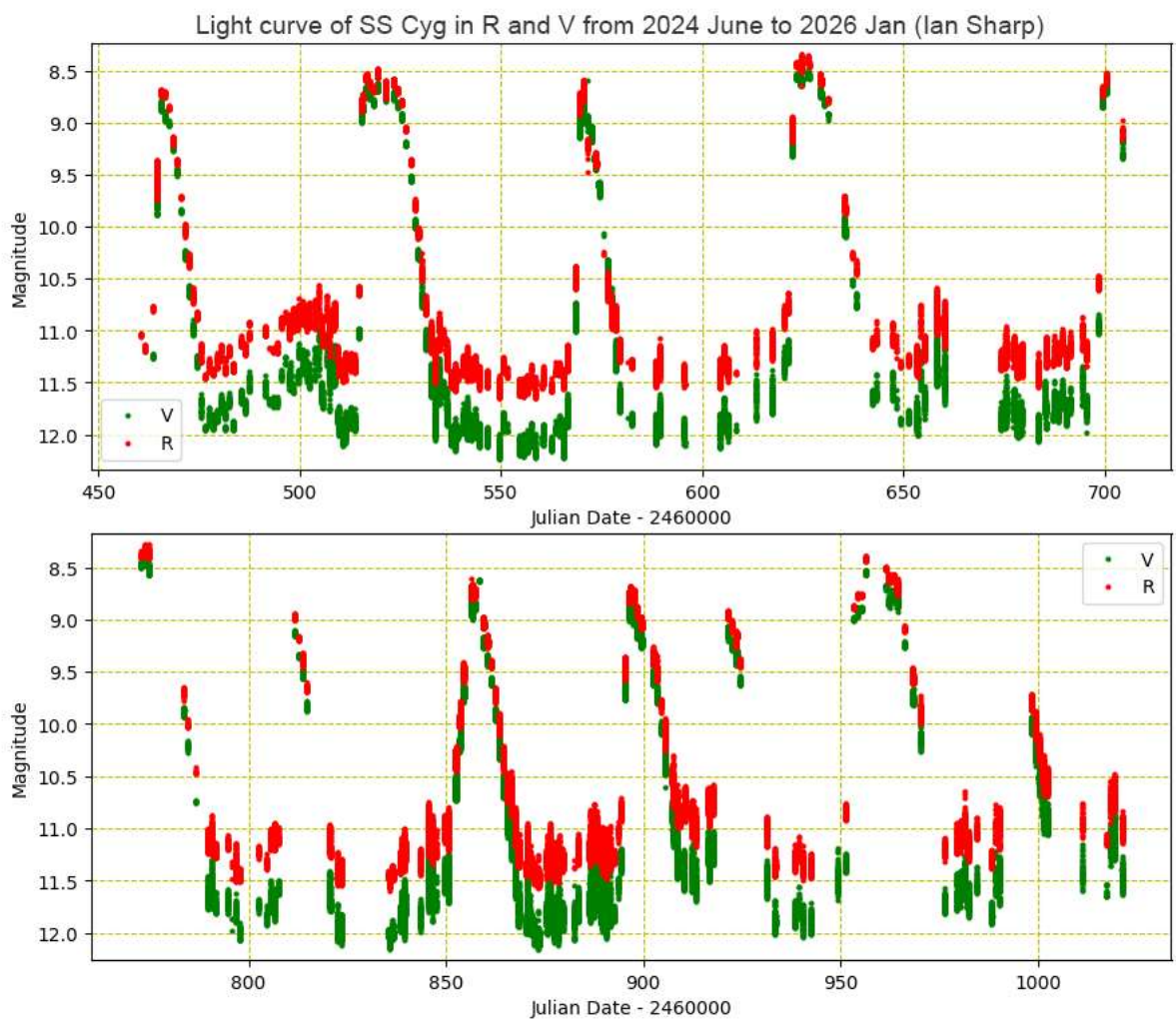

Figure 1. The light curve of SS Cygni using data from the author with R and V photometric filters from 2024 June to 2026 January. A total of 66,062 magnitude measurements are shown on the two charts in Figure 1 and the familiar outbursts are clearly shown. It can also be seen that the variability and differences in the V and R magnitudes are much greater in the quiescent phases.

To see more detailed structure and detect shorter-term variations in the light curve it is necessary to plot and analyse each night separately. The cadence of the exposures is approximately 45 seconds (including file download time) where the filter is swapped between R and V on every exposure so that the typical cadence between exposures of the same filter is around 90 seconds. Because of this it is not possible (due to the [Nyquist-Shannon sampling theorem](#)) to determine any periodic frequency components in the light curves with periods with of less than 180 seconds (3 minutes). Conversely the typical run length in any one night of around 2 to 4 hours restricts the lowest frequency component that can be extracted from the light curves.

The restrictions imposed by the minimum sampling frequency and the typical length of any nightly dataset described above determine the upper and lower frequencies of any periodic oscillations that can be analysed-out of the light curves and, in this study, the limits are set to 0.25 to 8 hours.

## Previous Studies

There are several studies of SS Cygni that specifically investigate periodic variations in brightness, particularly rapid oscillations and quasi-periodic oscillations seen during outbursts or at high time resolution. Below is a list of studies that include findings of oscillations in the light curve of SS Cyg.

1. **[Robinson & Nather (1979)](#)** – *Quasi-periodic luminosity variations in dwarf novae*.
   Includes SS Cyg and reports ~32 s quasi-periodic oscillations in its light curves.

2. **[Horne, Gomer, et al. (1980)](#)** – *Phase variability in the rapid optical oscillations of SS Cygni*.
   Includes high-speed optical photometry reporting ~8 s oscillations.

3. **[Hildebrand, Spillar & Stiening (1981)](#)** – *Observations of fast oscillations in SS Cygni*.
   Early detection of rapid photometric oscillations during outburst.

4. **[Cordova et al. (1984)](#)** – *Observations of quasi-coherent soft-X-ray oscillations in U Gem and SS Cyg*.
   Quasi-coherent X-ray oscillations in SS Cyg during decline from outburst.

5. **[Mauche & Robinson (2001)](#)** – *First simultaneous optical and EUV observations of the quasi-coherent oscillations of SS Cygni.*
   EUV and optical oscillations observed simultaneously; includes frequency changes ("frequency doubling").

6. **[Mauche (1996)](#)** – *Quasi-coherent oscillations in the extreme ultraviolet flux of SS Cygni.*
   Detailed study of EUV oscillation periods changing throughout outburst.

7. **[Mauche (2003 review)](#)** – *Optical, UV, and EUV oscillations of SS Cygni in outburst*.
   Summarizes multiwavelength quasi-periodic phenomena and places SS Cyg's oscillations in context with other compact binaries.

The studies listed above describe oscillations that are typically classified as dwarf nova oscillations (DNOs) or low-coherence quasi-periodic oscillations (QPOs), often with periods of ~7–40 s during outbursts. The studies above span optical, Extreme UV, and X-ray bands. DNOs are higher frequency (tens of seconds), more coherent oscillations linked to the white dwarf's rapidly spinning equatorial belt during outburst, while QPOs are lower frequency (hundreds of seconds), broader oscillations likely caused by vertical waves or thickening in the inner accretion disk.

For citations that include SS Cygni's quasi-periodic light curve behaviour during outburst in longer timescale studies or in photometric surveys, see **Voloshina et al. (2012)**. which discusses QPOs from ground-based light curves indicating quasi-periodic signals near peak and decline.

It appears that there are no studies that cover the frequency range with periods between 0.25 and 8 hours as discussed in this study. Additionally, it is clear from Figures 2 and 3 that there is more variability in magnitude during the quiescent phases than the outbursts and this becomes more apparent when the individual nightly light-curves are examined. However, most other studies seem to ignore the quiescent phases.

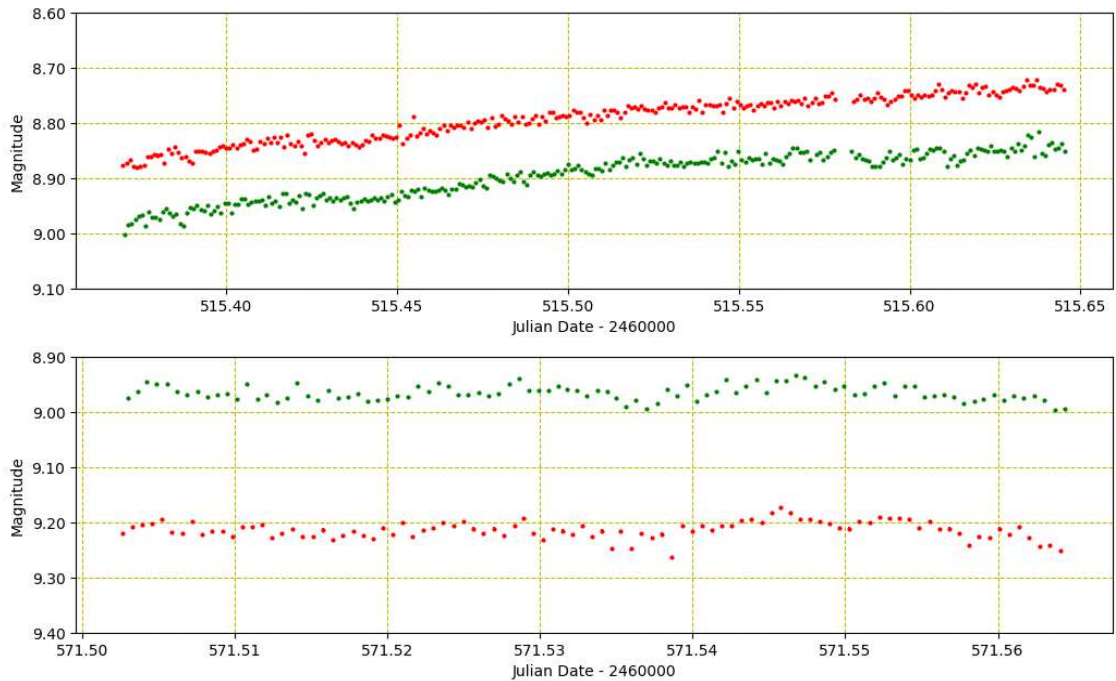

Figure 2. Two examples of nightly light curves in both R and V filters taken during outburst. Note the lack of variability compared to curves captured during the quiescent phases.

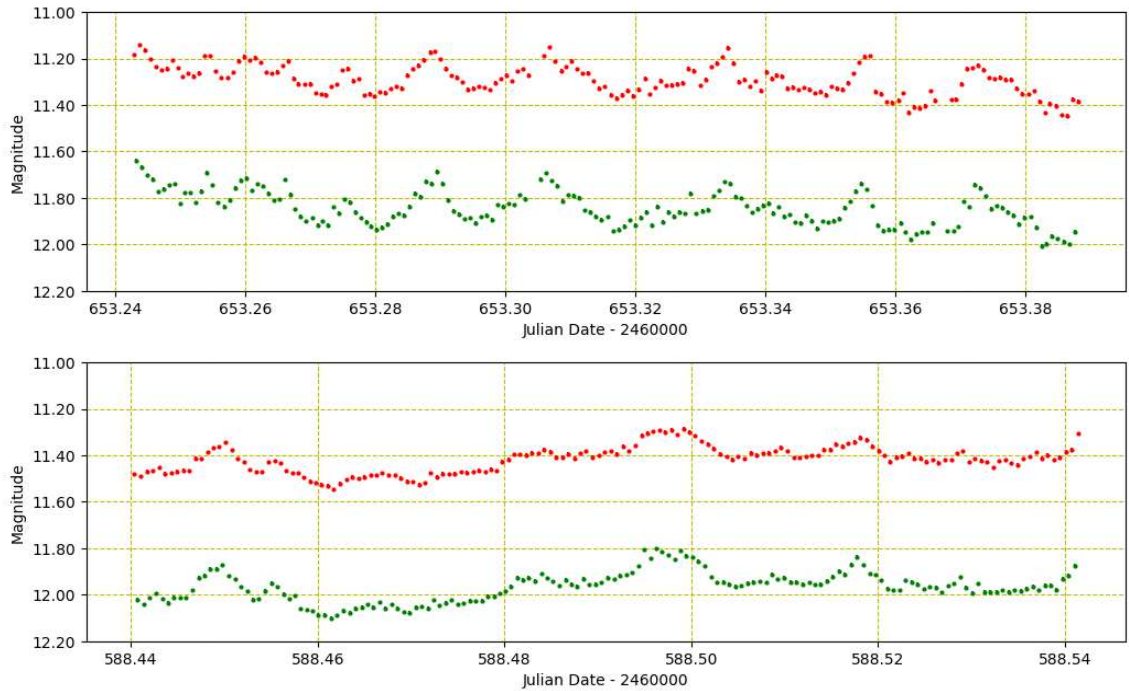

Figure 3. Two examples of nightly light curves in both R and V filters taken during quiescence. Note the variability in magnitude is much greater than the light curves captured during outburst.

### Data Analysis

The author used his own Python programmes to perform the power spectrum analysis using the Generalised Lomb-Scargle periodogram (GLS) which is a powerful technique for finding periodic signals in unevenly sampled data, creating a periodogram (power spectrum) by fitting sinusoids to data points. The Python library used to implement the GLS is part of the [PyAstronomy](PyAstronomy) packages.

The Generalised Lomb-Scargle periodogram was selected for this analysis because it is specifically designed to handle the irregular sampling patterns typical of ground-based astronomical observations. Unlike standard Fast Fourier Transform methods, which require evenly spaced data points, GLS can effectively analyse time series with gaps caused by weather, equipment issues, and meridian flips. Additionally, GLS does not require interpolation or gap-filling techniques that could introduce artificial periodicities. The algorithm is particularly well-suited to detecting weak periodic signals in the presence of noise, making it ideal for identifying the subtle quasi-periodic oscillations expected in SS Cygni's light curves. The ability to calculate meaningful False Alarm Probability (FAP) levels for irregular data further strengthens the statistical significance of detected periodicities, which is crucial when analysing the large number of individual nightly datasets in this study.

The workflow from raw data to the final GLS results is as follows. All Python scripts were written by the author:

1. Filter and download the data from the BAA photometric database into which the author has already uploaded all data on a nightly basis. This is a convenient way to obtain a comma-separated values (CSV) file of all the data rather than attempting to stitch together hundreds of separate csv files stored on the author's hard disk.

2. Run a custom Python script to read and process the csv file from step 1 to split the data into separate CSV files (one per night) with columns containing the Julian Date, Magnitude, Magnitude error and Filter. This is done by detecting the large gaps during daylight hours.
3. Run a custom Python script to plot each CSV data file for visual inspection. This script also rejects any files with too few data points and also files that have too many gaps. Gaps can be caused by clouds, meridian flips and various technical equipment issues. In this study datasets with less than 30 points are rejected as are datasets containing gaps over 15 minutes. Note that these rejection parameters can be easily varied in the code.
4. Using only the files that pass the tests in step 3, run each nightly file through a custom Generalised Lomb-Scargle algorithm (written in Python). This plots the results of the GLS and also stores the frequency (and period) of the peaks coming out of the power spectrum. The result is a single CSV file containing the details of all the GLS power peaks from all the files.
5. Lastly, plot a histogram of the results from step 4. This is also performed in Python code.

It should be noted that the above workflow produces over 300 light curve graphs (typically with both R and V curves) plus another set of 300+ associated power spectrum graphs arising from the GLS algorithm. Clearly, there are too many graphs to display in this paper and therefore a small sample will have to suffice and Figure 4 shows twelve examples. However, the author has uploaded all the graphs on a webpage at the following URL: http://python.astro-sharp.com/graphs.

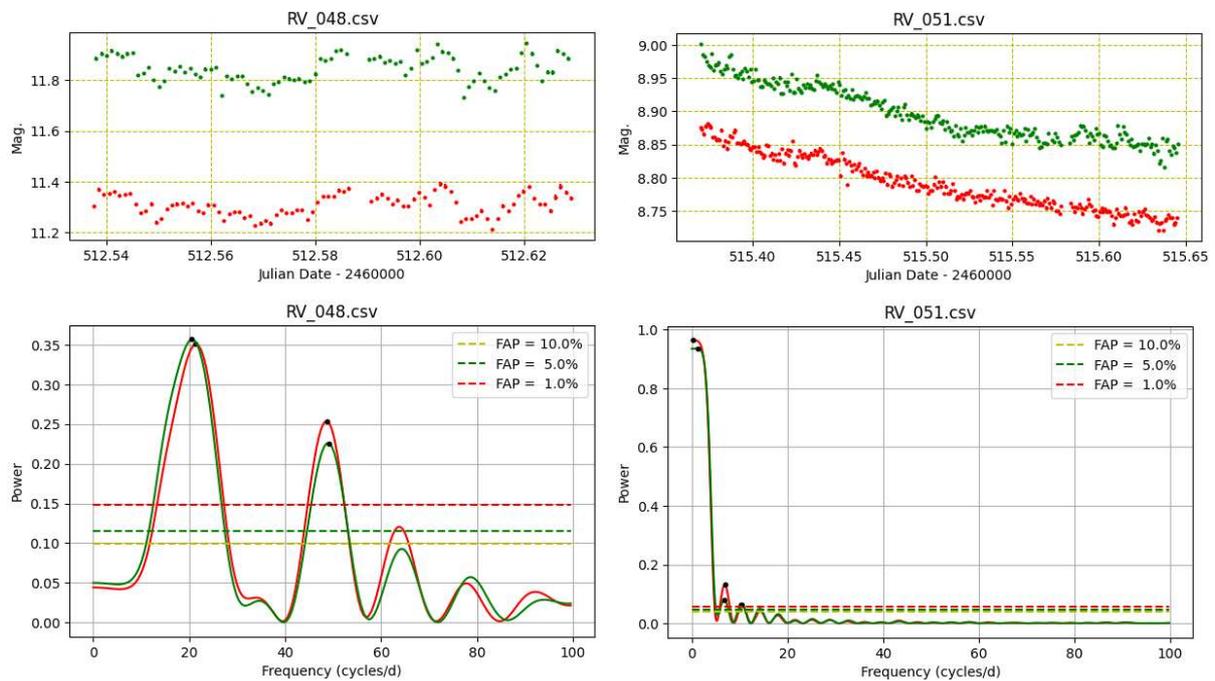

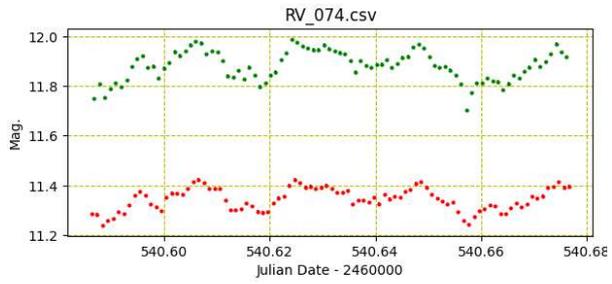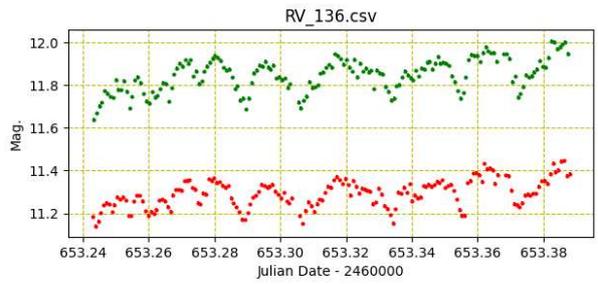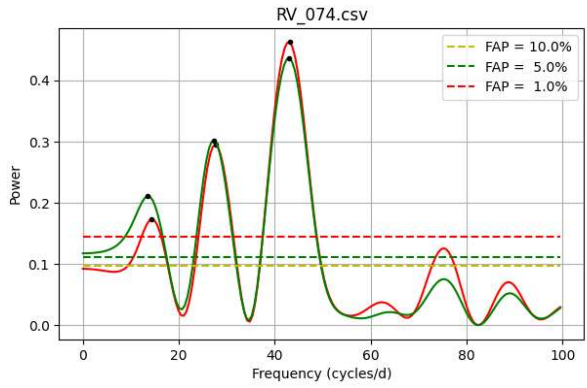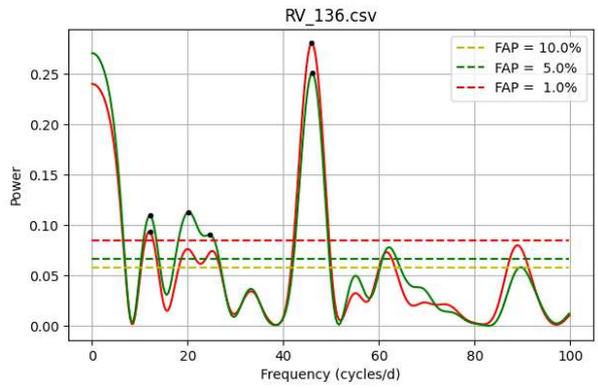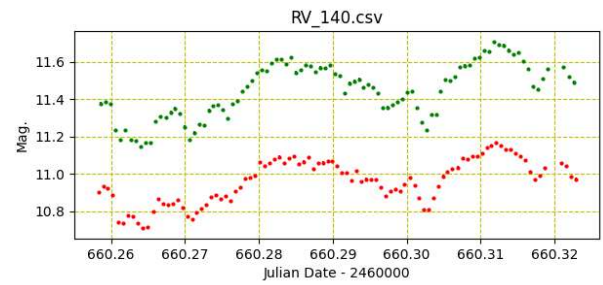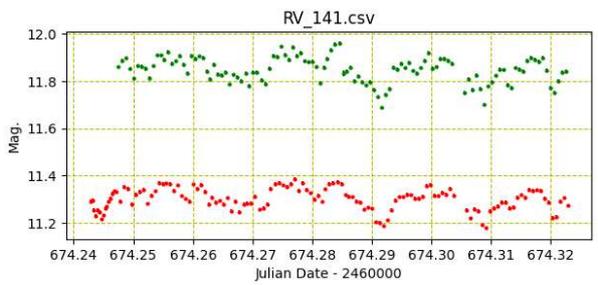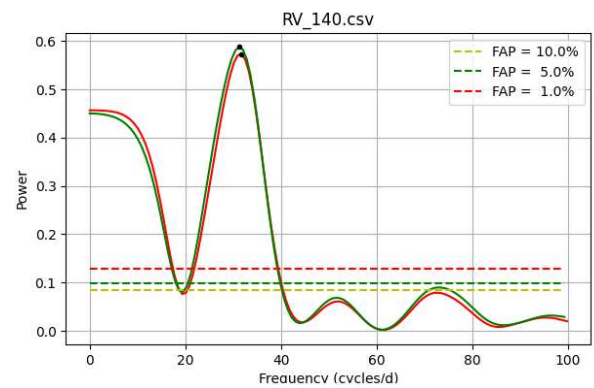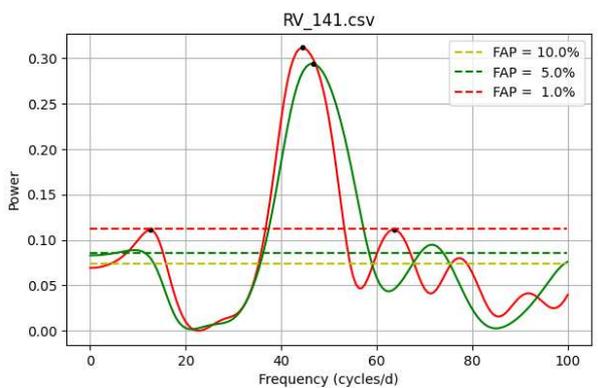

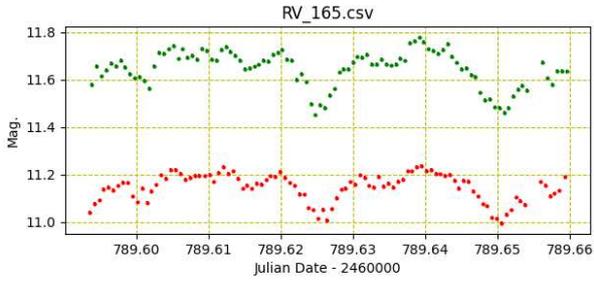
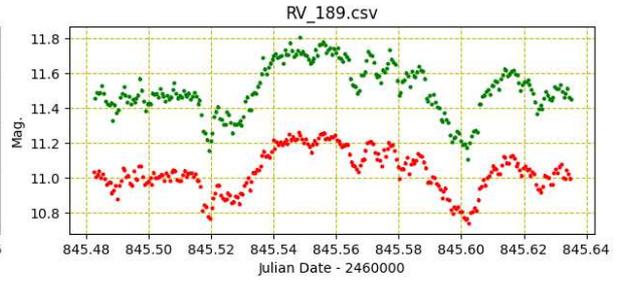
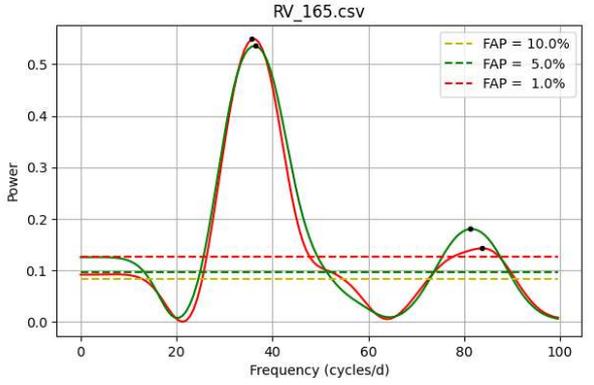
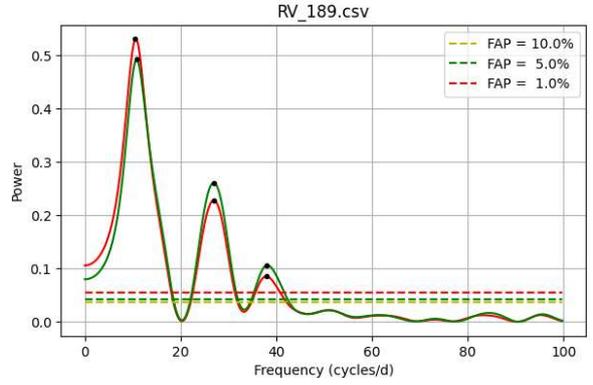
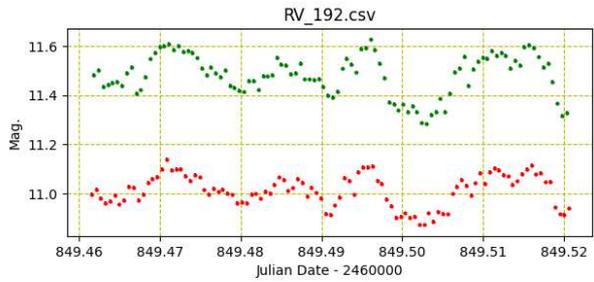
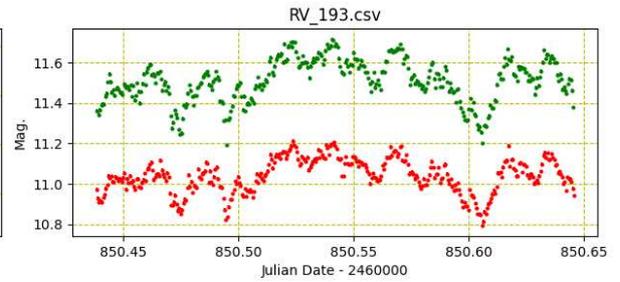
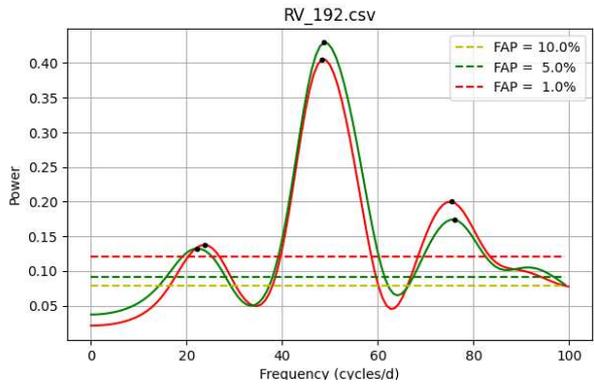
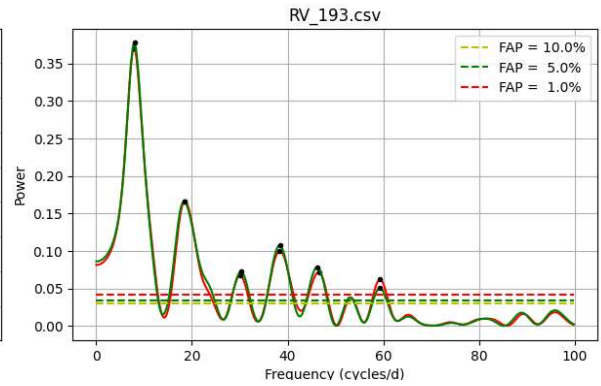

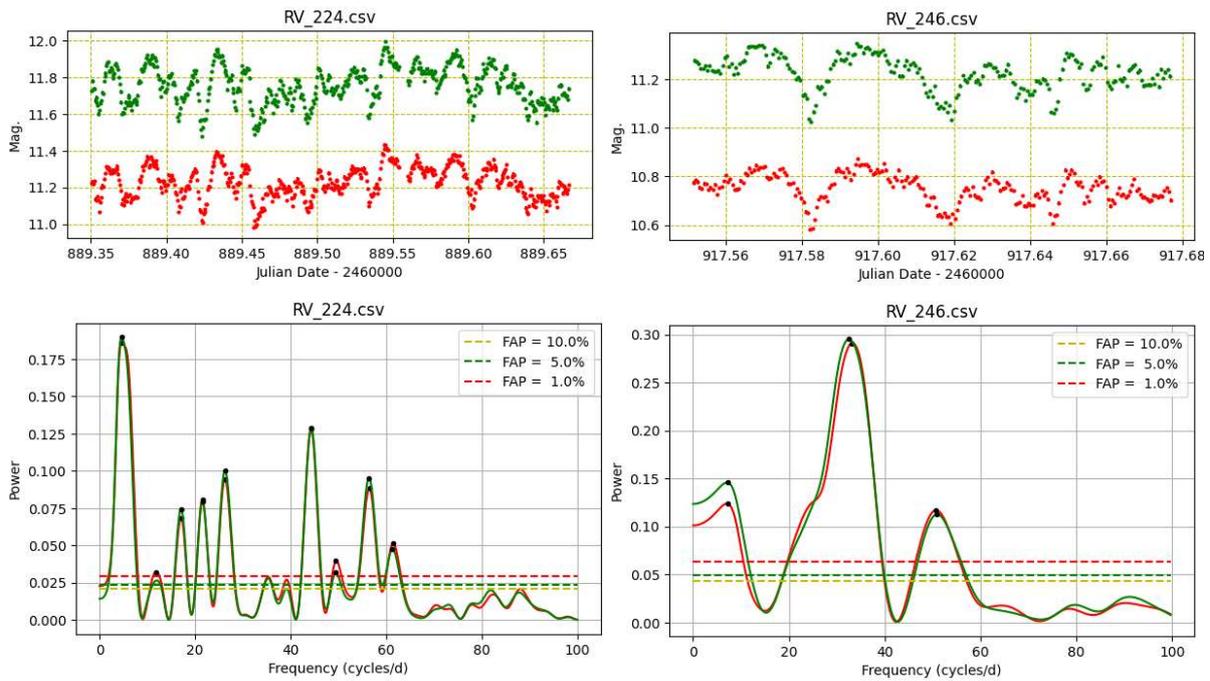

Figure 4. A sample of 12 light curves along with their associated power spectra from the GLS computations. The light curve is shown above the power spectrum for each result. Note that this is a small sample from more than 300 similar results. The titles of the graphs are the file names of the CSV files being processed.

The algorithm for the GLS written in Python has been very thoroughly tested by comparing the results to those produced by the very highly regarded Peranso software package. The author has compared dozens of randomly selected light curves and has always found the results to be identical to the eye. This includes the FAP levels; see Figure 5. for an example. The main driving force to develop a Python programme was to be able to batch-run the GLS algorithm on hundreds of light curves and to be able to output the hundreds of power spectrum peaks for further analysis. Additionally, the batch can be re-run with different parameters if required.

A low FAP indicates that a peak is unlikely to be generated by noise, representing a significant detection of a frequency component. It is defined as the probability that random noise would produce a peak as high or higher than the observed peak. In the following analysis, only peaks with a power greater than an FAP of 0.01 (1%) are used.

It is mostly the case that the light-curves and the power spectra produced for the R and V filters produce very similar results. The shapes of the light curves are very similar apart from having different absolute magnitude values (i.e., R brighter than V). The shapes of the curves and the peaks from the GLS are also strikingly similar.

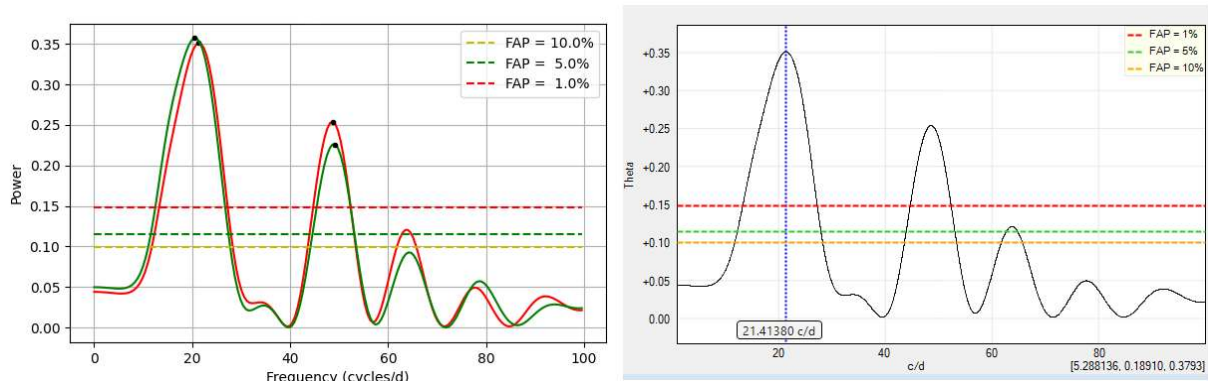

Figure 5. An example of a comparison of the GLS power spectrum results between the author's Python version and the Peranso software. Note that the Peranso graph is only showing the result for the R filter. The results are virtually identical including the False Alarm Probability levels.

The final step in the workflow (described above) is to plot the distribution of the detected peaks in the GLS power spectrum on a histogram graph. Figure 6 shows two histogram plots of periods where the first has a binning width of 5 minutes and the second with 10 minutes. Note that the histograms contain the combined frequency peaks from both the R and V data. The author experimented with separating the R and V results into separate histograms but the resulting plots were so similar that it seemed sensible to combine them. The x-axes of the histograms were extended beyond 4 hours to include the 6.6-hour binary period of SS Cyg which does not show up as a peak on the histogram.

Referring to the right-hand histogram in Figure 6 where the bin size of each bar is 10 minutes the top 4 bars are clustered together and contain 461 power spectrum peaks which is 55% of the total. The period range of the top 4 bars covers a span of 14.6 to 54.5 minutes where the centre of the highest bar is 30 minutes and has a count of 141 peaks.

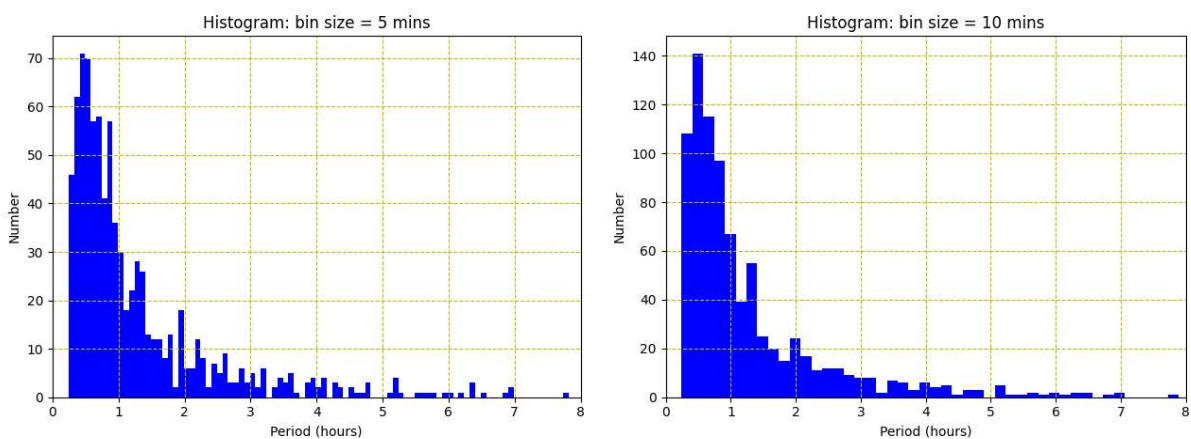

Figure 6. Two histograms showing the distribution of the periods of power spectrum peaks extracted from the data. The period in hours on the x-axis has been limited to 8 hours but the charts are showing 98% of the data and any peaks beyond 8 hours are insignificant.

## Summary and Conclusions

The above analysis of the periodic frequency content of the light curves of SS Cygni shows that there are many QPOs with periods in the range 0.25 to 8 hours (as dictated, in this study, by the sampling frequency due to the cadence of image acquisition). Most of the variability and the majority of the detected power spectrum peaks occur during the quiescent portions, not the outburst portions of SS Cygni's light curve. This would suggest that these really are QPOs and not DNOs which are typically associated with outbursts.

More than 55% of the power spectrum peaks are clustered together in the period range of approximately 15 to 60 minutes (96 to 24 cycles/day). The highest number of peaks in any 10-minute window occur centred on a period of 30 minutes (48 cycles/day). The author has not been able to find other studies that have detected QPOs in this range of periods, probably because the quiescent periods have been under-represented.

## Further Work

The author intends to carry on the regular monitoring of SS Cyg to confirm that the behaviour described above continues. During outburst it should be possible to halve the exposure time and thus look for QPOs or even DNOs at higher frequencies. Broadening the research with potential collaborations or complementary observations could be fruitful.

## Acknowledgments

Thanks to Jeremy Shears and Gary Poyner of the BAA Variable Star Section for their encouragement. ChatGPT was found to be very useful in finding other related studies.